\documentclass{camera}

\usepackage{graphicx} 

\begin{document}

\title{Commissioning of the Silicon Drift Detectors of the ALICE experiment at the LHC}
\author{Emanuele Biolcati \\ for the ALICE collaboration}
\organization{Universit\`a degli Studi e INFN di Torino}

\maketitle
  
  \begin{abstract}
  Silicon Drift Detectors (SDD) equip the two central layers of the Inner Tracking System of the ALICE experiment at the LHC. Main results of systematic studies of detector performance including noise, gain, drift speed and charge collection measurements will be reported.
  \end{abstract}

\section{ALICE experiment and SDD module}
Silicon Drift Detectors (SDD)~\cite{Beole:2007dc} offer two-dimensional position reconstruction with a limited number of read-out channels, high granularity and good multi-track detection capability. For these reasons, SDD have been chosen to equip two layers of the ALICE Inner Tracking System, the detector closest to the beam axis~\cite{PPR2}.

The SDD operating principle \cite{Biolcati} is based on the measurement of the time necessary for the electrons produced by the crossing particle to drift, in an adequate electrostatic field, from the generation point to the collecting anodes. The first coordinate (\emph{anodic}) is obtained from the centroid of the charge distribution along the anodes, while the distance of the crossing point from the anodes (\emph{drift coordinate}) is determined by the measure of the drift time and requires a precise knowledge of the drift speed.

\section{Calibration of the Silicon Drift Detectors}

\paragraph{Noise and gain} 
In 2008, 97.6\% of each SDD module have taken data. 
During a \emph{pedestal} run, baseline and noise are measured for each channel of the 260~SDD modules mounted in the ITS (in total 133k channels\footnote{The number is obtained by: 256 anodes $\times$ 2 drift sides $\times$ 260 modules.}). 
Using these values, the AMBRA chip \cite{Biolcati} allows to equalize the baselines for each group of 64~channels. The baseline values, measured before the equalization, are Gaussian distributed around 50~ADC counts. The noise distribution is centered at 2.3~ADC count, well within the design goals. A Minimum Ionising Particle (MIP) passing close to the anodes generates a signal peak (after the baseline subtraction) of 120~ADC counts, so the signal-to-noise ratio $S/N \simeq 120/2.3 \simeq 52$. 
 A second calibration run type is called \emph{pulser} and it is meant to measure the gain for each electronic channel. This is obtained by sending a signal, called \emph{test pulse}, to the pre-amplifier input. 
During the data taking period with cosmic rays, between July and October 2008, it has been possible to monitor these parameters as a function of time, to check the stability of the detector. 
All the baseline mean variations are below 1\%. The maximum noise fluctuations are 3\% during all the period. The RMS of the gain distribution is about 10\% of the mean. 

\begin{figure}[!ht]
  \centering
  \includegraphics[width=0.32\textwidth,height=4cm]{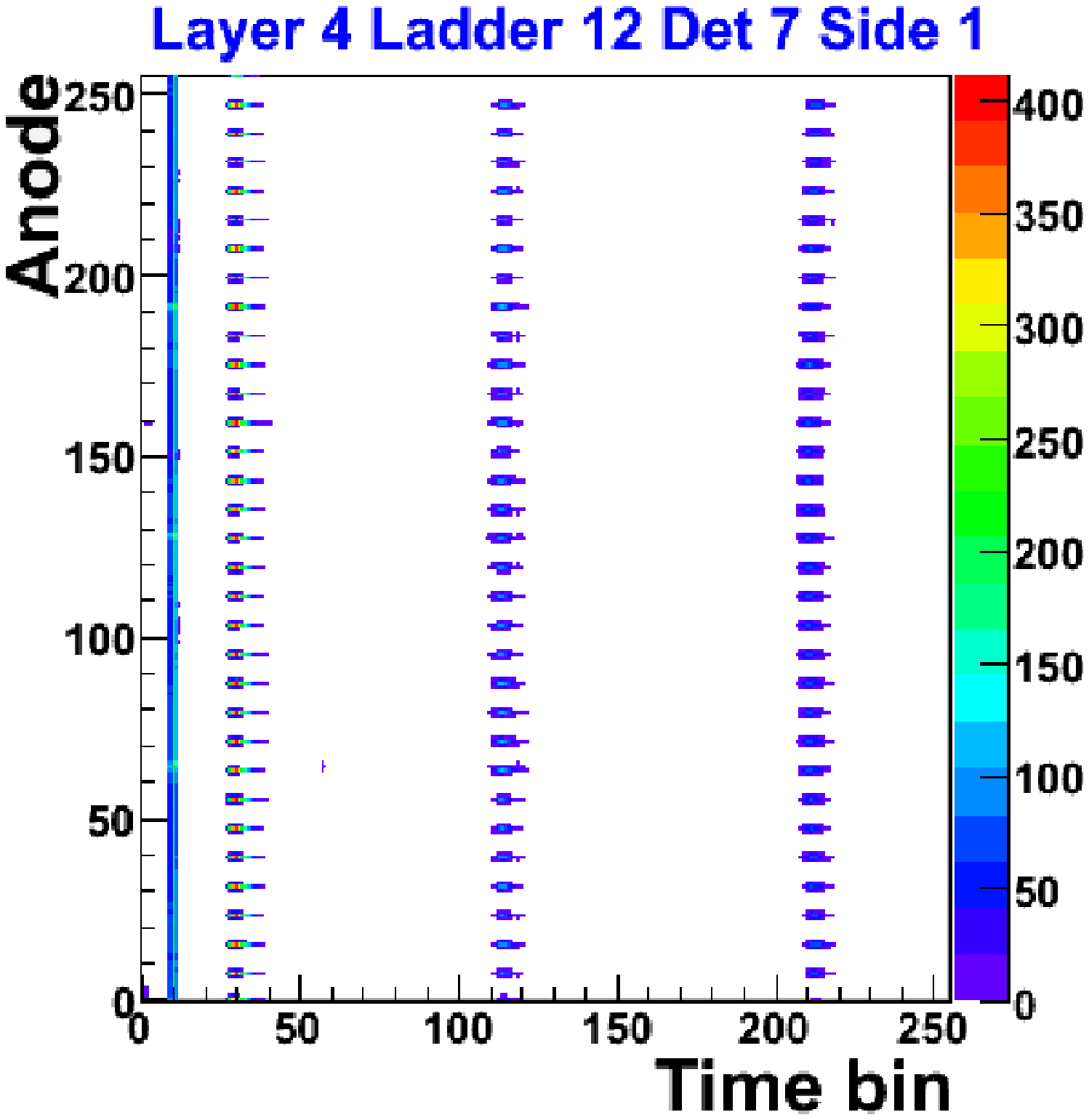}
  \includegraphics[width=0.32\textwidth,height=4cm]{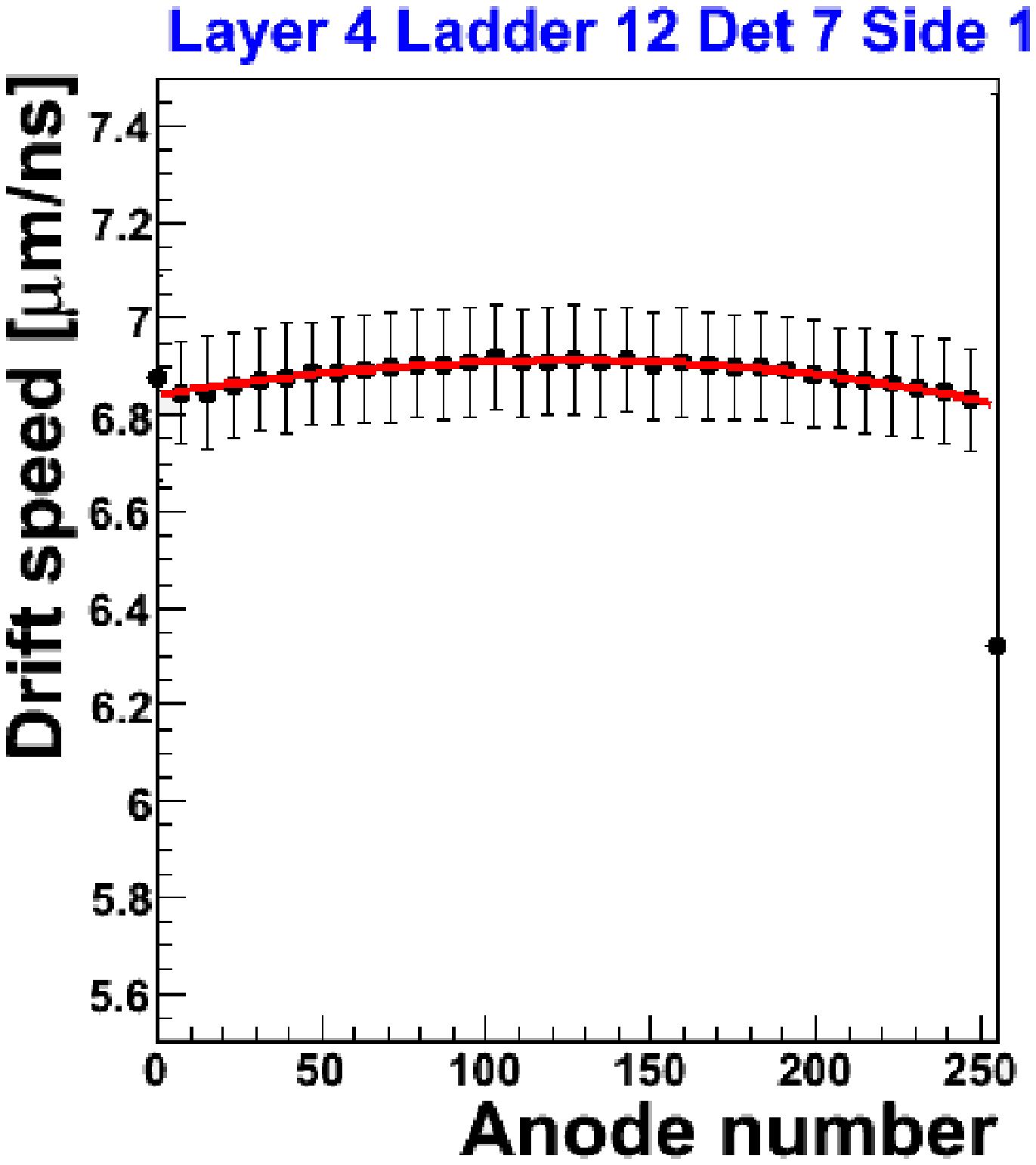}
  \includegraphics[width=0.32\textwidth,height=4cm]{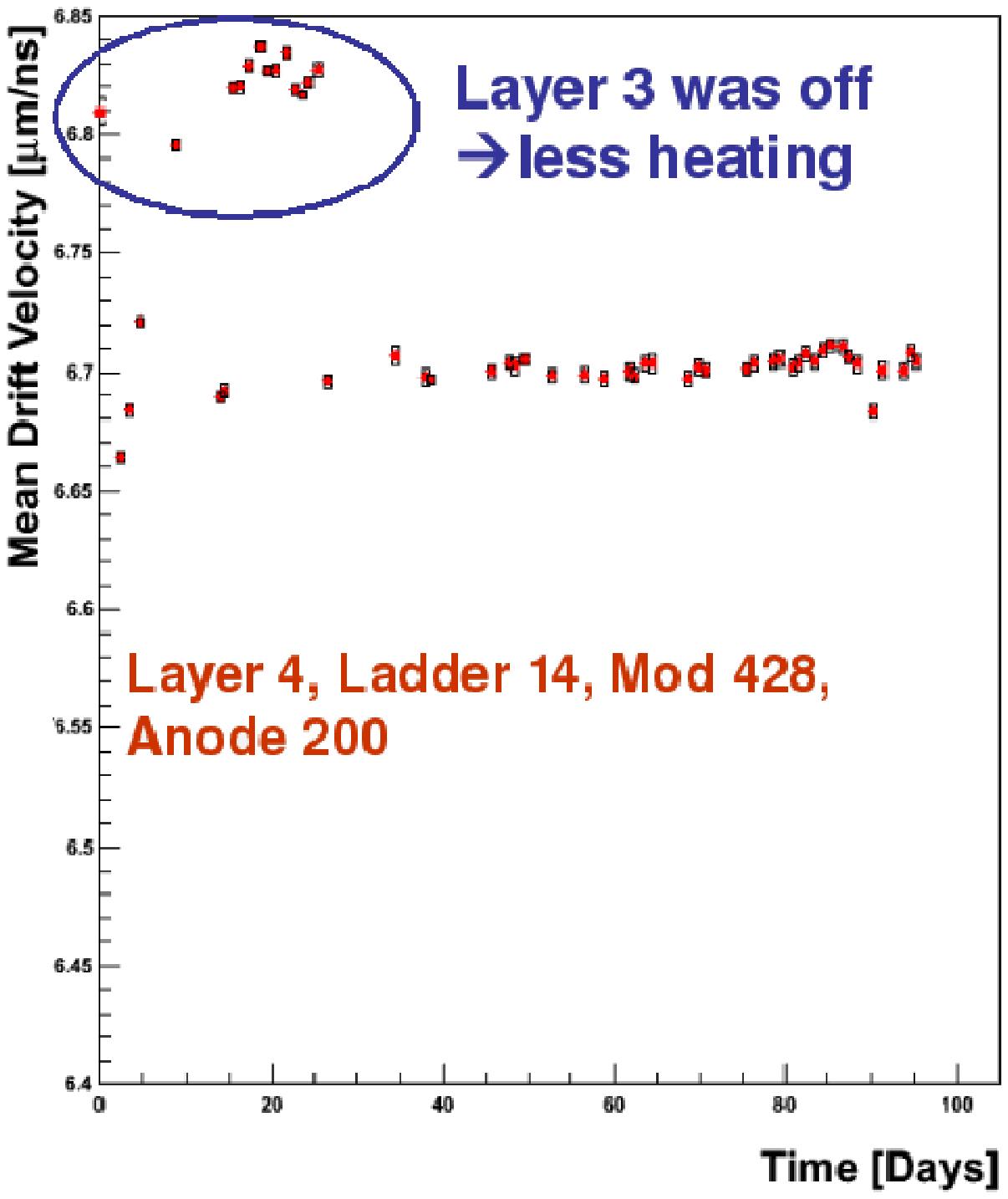}\\
  (a) \hspace{3.5cm} (b) \hspace{3.5cm} (c)
  \caption{\it (a) Injector rows. (b) Drift speed vs anode position. (c) Drift speed mean values vs time, for the outer SDD layer of ITS.}
  \label{driftspeed}
\end{figure}

\paragraph{Drift speed}
The design goal for the SDD resolution along $r\varphi$ (\emph{i.e.} drift coordinate) is 35~$\mu\mbox{m}$. A precise determination of the drift speed is therefore mandatory. This is complicated by the fact that the drift speed is strongly sensitive to temperature variations. A variation of 1~K changes the drift speed by 0.3~$\mu\mbox{m/ns}$ with a maximum systematic error of 1~mm on the drift position, because $\mu_{\mathrm{e}}\propto T(K)^{-2.4}$. Temperature gradients are present within each sensor because of the current circulating in the voltage divider implanted in each SDD module and must be corrected for. Moreover variations of the ambient temperature can take place even in the presence of the cooling system. 
It is therefore important to measure frequently the drift speed value in many positions on the detector. For this purpose three rows of 33 injectors consisting of MOS capacitors~\cite{Biolcati} are implanted in the sensitive area of each SDD drift region. From the fit of the 3 points (one per row) it is possible to calculate the drift speed values in 33 positions along the anodes. An \emph{injector} run aimed at measuring the drift speeds is performed every 8-10 hours during the data taking period.
In fig.~\ref{driftspeed}a the result from such a run is shown. The charge injected by the three lines is clearly visible because the zero-suppression alghoritm is active and all the cells below thresholds are not stored in the histogram. In the fig.~\ref{driftspeed}b the respective calculated drift speed values are shown. The anode dependence is caused by the temperature gradient, induced by the heat produced by the voltage divider implanted at the edges of the sensitive area, close to anodes 0 and 255. 
The difference between maximum and minimum values of drift speed is of  the order of 1\%. Data are fitted by a third degree polynomial and the fit parameters are stored in the Offline Condition Database of the ALICE experiment.
A difference of about 4\% between the two SDD layers is present, due to their different temperature. This is illustrated infig.~\ref{driftspeed}c where the drift speed values as a function of time for one anode are shown.
When the inner layer was off, the drift speed values in the outer layer were higher than the nominal ones by $\sim$2\%, due to the lower operating temperature.
During the cosmic data period (4 months) the drift speed variations on each anode were on average below 2~per mille. 

\begin{figure}[!ht]
  \centering
  \includegraphics[width=\textwidth]{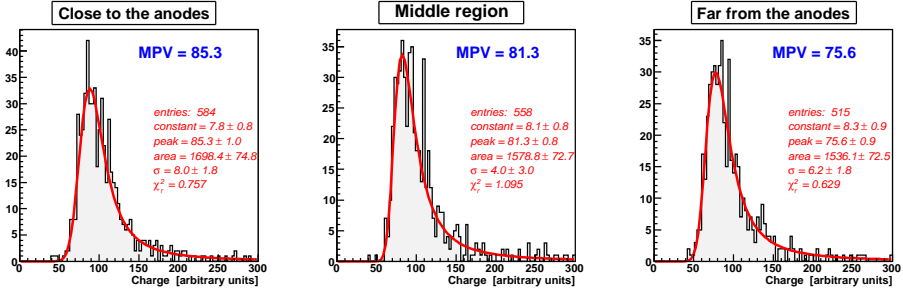}
  \caption{\it Cluster charge distribution vs drift distance.}
  \label{drift_dependence}
\end{figure}

\paragraph{Charge collection}
The charge produced by an atmospheric muon crossing the SDD sensor is collected in a group (\emph{cluster}) of contiguous anode/time bin cells above the threshold for zero-suppression. For each \emph{cluster}, it is possible to calculate the energy deposited in the detector volume by the atmospheric muon.
The energy distribution of about 18.000 \emph{clusters} is plotted in fig.~\ref{drift_dependence} and fitted by a convolution of a Landau and a Gaussian. 

From the fit it is possible to extract the Most Probable Value (MPV) and obtain a conversion factor from the detector response in ADC counts to physical units (keV), based on the known MPV for energy deposit of a MIP in 300~$\mu$m of Silicon. The MPV decreases with increasing drift length.
Electron clouds generated far from the collection anodes need more time to drift with respect to one generated close to the anodes\footnote{The maximum drift time at the nominal electric field of 500~V/cm is 5.3~$\mu s$.} and therefore they are spread out by Coulomb repulsion and diffusion effects \cite{Nouais}. 
The decrease of the MPV with increasing of drift time comes from the fact that the tails of the clusters are cut by the zero-suppression alghoritm. 
The difference is about 11\% and is properly accounted for by an offline correction. This correction is based on Monte Carlo (MC) simulations of the detector response. 

\paragraph{Conclusions}
During the 2008 runs with cosmic rays, it has been possible to calibrate the Silicon Drift Detectors of the ALICE Inner Tracking System.

The front-end electronics parameters (noise and gain) and the drift speed have been measured frequently via dedicated calibration runs. They are stable and all the values are stored in the Offline Condition Database of the experiment.
As far as charge collection is concerned, the ADC-to-keV conversion factor has been tuned and the MC based correction for the charge dependence on drift distance has been evaluated.


\begin{thebibliography}{99}

\bibitem{Beole:2007dc} S.~Beole {\it et al.},
  Nucl.\ Instrum.\ Meth.\  A {\bf 570} (2007) 236.

\bibitem{PPR2} B.~Alessandro {\it et al.}  [ALICE Collaboration],
  J.\ Phys.\ G {\bf 32} (2006) 1295.  
 
 \bibitem{Biolcati} G.~Batigne {\it et al.},
  JINST {\bf 3}, P06004 (2008).

\bibitem{Nouais}
  D.~Nouais {\it et al.},
  Nucl.\ Instrum.\ Meth.\  A {\bf 477}, 99 (2002).
  
\end{thebibliography}
\end{document}